\documentclass[10pt,letterpaper]{article}
\usepackage[top=0.85in,left=2.75in,footskip=0.75in]{geometry}

\usepackage{amsmath,amssymb}

\usepackage{changepage}

\usepackage{textcomp,marvosym}

\usepackage{cite}

\usepackage{nameref,hyperref}

\usepackage[right]{lineno}

\usepackage[nopatch=eqnum]{microtype}
\DisableLigatures[f]{encoding = *, family = * }

\usepackage[table]{xcolor}

\usepackage{array}

\usepackage{float}
\usepackage{multirow}
\definecolor{blanco}{rgb}{1,1,1}
\usepackage{color}
\usepackage{tabularray}
\UseTblrLibrary{diagbox}
\usepackage{transparent}
\usepackage{rotating}
\usepackage{tabularray}
\usepackage{makecell}
\usepackage{colortbl}
\usepackage{hhline}
\usepackage{bm}
\usepackage{units}

\newcolumntype{+}{!{\vrule width 2pt}}

\newlength\savedwidth

\raggedright
\usepackage[aboveskip=1pt,labelfont=bf,labelsep=period,justification=raggedright,singlelinecheck=off]{caption}

\makeatletter
\renewcommand{\@biblabel}[1]{\quad#1.}
\makeatother

\usepackage{lastpage,fancyhdr,graphicx}
\usepackage{epstopdf}
\fancyheadoffset[L]{2.25in}
\fancyfootoffset[L]{2.25in}
\begin{document}
\vspace*{0.2in}

\begin{flushleft}
{\Large
\textbf\newline{Geometric immunosuppression in CAR-T cell treatment: Insights from mathematical modeling} % Please use "sentence case" for title and headings (capitalize only the first word in a title (or heading), the first word in a subtitle (or subheading), and any proper nouns).
}
\newline
\\
Silvia Bordel-Vozmediano\textsuperscript{1,2,*},
Soukaina Sabir\textsuperscript{1,2},
Lucía Benito-Barca\textsuperscript{3},
Bettina Weigelin\textsuperscript{4,5},
Víctor M. Pérez-García\textsuperscript{1,2}.
\\
\bigskip

\textbf{1} Mathematical Oncology Laboratory (MOLAB), Instituto de Matemática Aplicada a la Ciencia y la Ingeniería, Universidad de Castilla-La Mancha, 13005 Ciudad Real, Spain
\\
\textbf{2} Departamento de Matemáticas, Escuela Técnica Superior de Ingenieros Industriales, Universidad de Castilla-La Mancha, 13005 Ciudad Real, Spain
\\
\textbf{3} Facultad de Ciencias Experimentales, Universidad Francisco de Vitoria, 28223 Pozuelo de Alarcón, Spain.
\\
\textbf{4} Department of Preclinical Imaging and Radiopharmacy, Multiscale Immunoimaging, Eberhard Karls University, Tübingen, Germany
\\
\textbf{5} Cluster of Excellence iFIT (EXC 2180) ``Image-Guided and Functionally Instructed Tumor Therapies'', Eberhard Karls University, Tübingen, Germany
\\
\bigskip

% Insert additional author notes using the symbols described below. Insert symbol callouts after author names as necessary.
% 
% Remove or comment out the author notes below if they aren't used.
%
% Primary Equal Contribution Note
%\Yinyang These authors contributed equally to this work.

% Additional Equal Contribution Note
% Also use this double-dagger symbol for special authorship notes, such as senior authorship.
%\ddag These authors also contributed equally to this work.

%\textcurrency Current Address: Dept/Program/Center, Institution Name, City, State, Country % change symbol to "\textcurrency a" if more than one current address note
% \textcurrency b Insert second current address 
% \textcurrency c Insert third current address


% Group/Consortium Author Note
%\textpilcrow Membership list can be found in the Acknowledgments section.

% Use the asterisk to denote corresponding authorship and provide email address in note below.
*silvia.bordel@uclm.es

\end{flushleft}
\section*{Abstract}

Chimeric antigen receptor T (CAR-T) cell therapy has emerged as a promising treatment for hematological malignancies, offering a targeted approach to cancer treatment. Understanding the complexities of CAR-T cell therapy within solid tumors poses challenges due to the intricate interactions within the tumor microenvironment. Mathematical modeling may serve as a valuable tool to unravel the dynamics of CAR-T cell therapy and improve its effectiveness in solid tumors. This study aimed to investigate the impact of spatial aspects in CAR-T therapy of solid tumors, utilizing cellular automata for modeling purposes.
Our main objective was to deepen our understanding of treatment effects by analyzing scenarios with different spatial distributions and varying the initial quantities of tumor and CAR-T cells. Tumor geometry significantly influenced treatment efficacy in-silico, with notable differences observed between tumors with block-like arrangements and those with sparse cell distributions, leading to the concept of immune suppression due to geometrical effects. This research delves into the intricate relationship between spatial dynamics and the effectiveness of CAR-T therapy in solid tumors, highlighting the relevance of tumor geometry in the outcome of cellular immunotherapy treatments. Our results provide a basis for improving the efficacy of CAR-T cell treatments by combining them with other ones reducing the density of compact tumor areas and thus opening access ways for tumor killing T-cells.

% Use "Eq" instead of "Equation" for equation citations.
\section*{Introduction}The advent of immunotherapy is revolutionizing cancer treatment. Among the many different immunotherapies available today, cellular immunotherapies based on modified patient-derived immune cells have emerged as promising concept  \cite{Waldman}.
Chimeric antigen receptor (CAR) T-cell therapy, one example of cell-based immunotherapies, relies on genetically engineering immune cells to express a synthetic receptor that targets surface antigens on tumor cells. CARs are composed of an antibody-derived extracellular domain that confers high specificity for tumor antigens, linked to intracellular signaling domains that drive T cell activation and effector function upon antigen binding. The generation of CAR-T cells begins by isolating T cells from the patient’s blood, followed by ex vivo activation, expansion and genetic modification to express the CAR receptor. Once reinfused  into the patient’s bloodstream, CAR-T cells home to lymphatic organs and tumor tissue where they selectively bind to cancer cells expressing the targeted antigen. Upon binding, the CAR receptor triggers the release of cytotoxic vesicles, leading to the destruction of the malignant cells \cite{June,Sterner}.

Initially celebrated for its remarkable success in treating hematological malignancies  \cite{Zhang,Lu}, CAR-T cell therapy has faced challenges in addressing solid tumors. They are highly heterogeneous, including diverse antigen expression that limits tumor recognition and highly variable tumor microenvironments (TMEs) that inhibit CAR-T cell effector functions \cite{Uslu}. Despite these challenges, recent advancements in CAR-T cell engineering combined with a deeper understanding of tumor biology and immunology have renewed optimism for CAR-T cell success in solid tumors   \cite{Albelda,Guzman}. CAR-T cells have e.g., been genetically modified to enhance their effector functions, prolong their survival and improve their migratory capacity within the TME \cite{Schmidts}. However, the complexity and redundancy of immunosuppressive mechanisms still make it challenging to identify which modifications are critical for optimizing CAR-T cell performance in the TME. 

Mathematical models, designed to describe, quantify, and predict multifaceted cell behavior, offer great potential for enhancing our understanding of CAR-T cell dynamics in solid tumors. By integrating experimental data and computational simulations, mathematical models may provide insights into the mechanisms underlying CAR-T cell efficacy or failure in the solid tumor microenvironment. Modelling can thus help unravel the interactions between tumor and immune cells in relation to microenvironmental parameters. Consequently, we have witnessed an extensive utilization of mathematical approaches in this domain in recent years. In hematological cancers, such as leukemia and lymphoma, mathematical models have focused on describing the kinetics of CAR-T cell expansion, interaction with tumor cells and the immune microenvironment within the bloodstream and lymphatic system and treatment associated toxicity  \cite{Liu, Ode, salvi, Mo, victor, roe, barros, ivana, alt}. Models used incorporated parameters such as CAR-T cell proliferation rates, tumor burden, cytokine levels, and immune cell interactions to predict treatment efficacy and potential adverse effects. However, most of those works have developed models of compartmental type and thus assumed implicitly a well mixed population of CAR-T and tumor cells which does not reflect the biology of solid tumors. Mathematical modeling of CAR-T cell function in solid tumors presents distinct challenges and models must encompass the varying spatial distribution of tumor cells and CAR-T cells within the tumor, distinct antigen expression levels, and diverse immunosuppressive TME niches. As models available to date have relied on assuming homogeneously mixed cancer and effector cell populations  (see e.g. \cite{brain, Russel, juan, Sahoo, Owens}), there remains a notable gap in the literature concerning the spatial dynamics of treatment response.

Our study aims to fill this gap by investigating the spatial dynamics of CAR-T cell therapy using cellular automata  \cite{CA,CA1}.  In recent years, the number of papers using CAs to study tumor growth accounting for immune interactions has been increasing  \cite{ger,mallet,pata,mallet1}. As example, Mallet et al. \cite{mallet1} developed a moderately complex, hybrid cellular automata model combined with a deterministic partial differential equation (PDE). In this model, they investigated the dynamics of tumor–immune system interactions by incorporating natural killer cells and cytotoxic T lymphocytes. They considered immune cell migration, death, and tumor lysis mediated by immune cells. Their findings revealed that depending on the strength of T cell recruitment and T cell killing kinetics, tumors exhibited stable or unstable oscillations, and in some cases, were completely eradicated. Zouhri et al. \cite{zouhri} improved this model by incorporating additional biological aspects such as the effects of interleukin-2 on the immune response.

In our study, we used a CA to investigate how the spatial distribution of tumor cells in the TME modulates tumor cell accessibility and consequently impacts CAR-T cell efficacy. We find that the spatial arrangement of tumor cells significantly impacts T cell expansion and T cell-mediated tumor elimination.  Our study provides a basis for mathematical models which inform which parameter of CAR-T cell function must be modulated in order to enable effective targeting of solid tumors.

\section*{Materials and methods}

The primary aim of this study was to explore how different tumor geometries influence the response to CAR-T cell treatment by simulating the interaction between tumor cells and tumor-specific CAR-T cells using cellular automata. We aimed to determine whether tumor cells are equally accessible to CAR-T cells across various geometries and to identify differences in treatment response caused by tumor cell distribution. Achieving this implementation requires careful consideration of the biological rules governing each cell type under investigation. Key processes to be incorporated include tumor cell proliferation and apoptosis thresholds as well as the viability and proliferative activity of CAR-T cells and the sustainability of their effector function. The details of all biological rules included in our CA are provided in this section. 

\subsection*{Physical structure}

The model consisted of a two-dimensional cellular automata operating on a square lattice with dimensions $n \times n$. Most of the simulations were run with $n=200$, thus a $4\times 10^4$ cell-lattice. Some studies were repeated with $n=400$ ($1.6 \times 10^5$ cells) to test the robustness of the results for larger lattice sizes, although always with similar results. Each site, can adopt one of three distinct states: 0 indicates an empty cell, 1 signifies a cell containing a CAR-T cell, and 2 denotes a cell occupied by a tumor cell. It is a biological fact that CAR-T cells are considerably smaller than typical tumor cells, i.e. about four times smaller in area. To account for this size difference in the cellular automata, we assumed that each CAR-T cell occupies one grid site, while each tumor cell occupied a larger area of four adjacent grid sites arranged in a square shape. Taking an average diameter of around 10 $\mu m$ for CAR-T cells, our spatial domain represents a tissue surface area of 4 $mm^2$.

In the model, we used a Moore neighborhood with a radius of 1. It is important to note that tumor cells exhibit a higher degree of neighboring connections compared to CAR-T cells. Each CAR-T cell is surrounded by 8 neighboring cells, while each tumor cell is adjacent to 12 neighboring cells.

In our study we considered two distinct geometries of the distribution of tumor cells within the lattice: dense  or sparse. Tumor cells arranged in a sparse manner were randomly distributed throughout the lattice, although avoiding physical overlappings, meaning that they were scattered uniformly without following a specific clustering pattern. In the dense tumor geometry, tumor cells were clustered in a compact mass located at the center of the lattice, forming a dense and cohesive structure. CAR-T cells were randomly positioned on the empty lattice sites in both scenarios. 

%\begin{figure}[H]
 %   \centering
  %  \includegraphics[width=1\columnwidth]{geometria.png}
   % \caption{\textbf{Comparaison of two potential tumor geometries. A.} Tumor cells sparse across the automaton lattice. \textbf{B.} Tumor cells arranged in a block formation at the lattice center.}
    %\label{geometria}
%\end{figure}

\subsection*{Transition rules and biological parameter choices}

Throughout the simulations, the status of individual lattice cells undergoes dynamic changes based on a set of transition rules governing the behavior and interactions of cells within the system. At each iteration of the simulation, these rules are applied, driving the ongoing updates to cell states and shaping the overall progression of the model.

\textbf{Temporal variables.} In our simulation framework, each discrete time increment corresponds to one hour. For scenarios with sparse tumor cells, we allotted a total duration of 1200 hours, reflecting a time span of 50 days. Simulations portraying tumor cells consolidated into a dense, clustered structure spanned 2160 hours, corresponding to a time frame of 90 days. 

\textbf{Tumor cell proliferation.} The proliferation parameter chosen was  $ \rho = 1/1000$ hours$^{-1}$ that corresponds to tumor doubling times over one month, thus describing fast-growing cancers are currently used only in late stages of the disease after  standard treatments have failed. Thus, in current clinical settings CAR-T cells are used against aggressive advanced forms of the disease justifying our choice  \cite{victor}. 

\textbf{Lifespan and proliferation of CAR-T cells.} Isolated CAR-T cells have a fixed lifespan ($\tau$) which we estimated  to be  $\tau=336$ hours (two weeks) based on recent literature  \cite{ghorashian,westera}. After the exhaustion of CAR-T cell, they are removed from the lattice of the automata since they do not play a significant role anymore.

\begin{figure}[H]
    \centering
    \includegraphics[width=1\columnwidth]{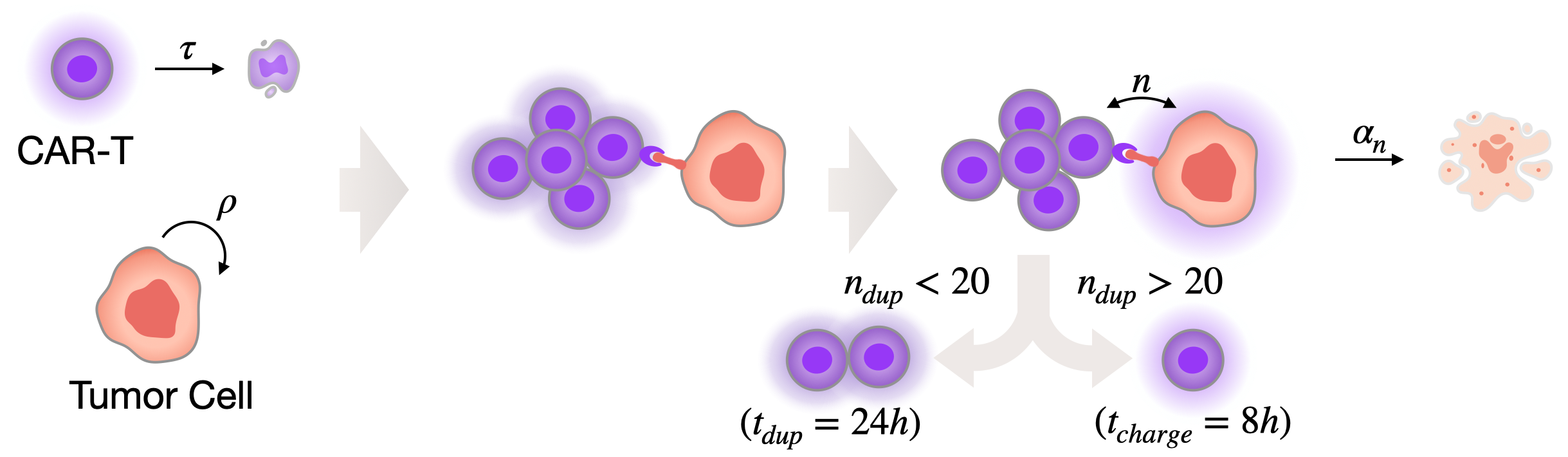}
    \caption{\textbf{Biological processes incorporated to the mathematical model.}  Isolated CAR-T cells have a fixed lifespan, $\tau$, while tumor cells proliferate at a fixed rate $\rho$. CAR-T cells are allowed to move through the lattice (but not tumor cells). When a CAR-T cell encounters a tumor cell, attaches to it and delivers its cytotoxic load. Tumor cells are capable of accumulating the cytotoxic load from multiple CAR-T cells. The probability of tumor cell death, denoted as $\alpha_n$, depends on $n$, which represents the number of CAR-T cells it has interacted with. After interactions between a tumor cell and CAR-T cells, two options arise depending on the value of the duplication counter, $n_{dup}$. If $n_{dup}<20$, CAR-T cells that have released their cytotoxic load enter a duplication phase lasting 24 hours ($t_{dup} = \unit[24]{h}$). Upon exiting this phase, CAR-T cells duplicate, with both the mother and daughter cells increasing $n_{dup}$ by 1. If $n_{dup} > 20$, it means that CAR-T cells that have released their cytotoxic load can no longer duplicate because they have reached their maximum capacity. In this case, the CAR-T cell remains stationary, awaiting a recharge of its cytotoxic load, which takes $t_{charge}=\unit[8]{h}$.}
    \label{esquema}
\end{figure}

\textbf{CAR-T – tumor cell interaction and killing kinetics.} The T-cell/target cell interaction parameters were obtained from recent experimental works addressing the biological details of those processes \cite{weigelin1,weigelin}. CAR-T cells possess a cytotoxic load that triggers apoptosis in tumor cells. Initially, each CAR-T cell has the maximum possible cytotoxic load. Encountering a tumor cell induces the release of the CAR-T cell’s cytotoxic load.  After each hit, CAR-T cells require 8 hours to fully replenish their cytotoxic load. Thus, throughout their lifespan, each CAR-T cell can execute a maximum of 42 cytotoxic hits. Nevertheless, it is plausible that during the simulation, some CAR-T cells may not come into contact with any tumor cells and, consequently, die without executing any targets. 

Each tumor cell can be targeted by multiple CAR-T cells simultaneously or sequentially, leading to the accumulation of cytotoxic load of each contact. As the number of contacts increase, so does the likelihood of the tumor cell’s demise, where the probability of death as a function of the number of contacts is denoted by $\alpha_n$. A single CAR-T cell contact results in a 5\% probability of cell death. However, with two CAR-T cells interacting simultaneously, the probability rises to 12\%. For three or four CAR-T cells, the likelihood of target cell death reaches 50\%. When a tumor cell is hit by five or six CAR-T cells simultaneously, the probability of death jumps to 80\%. Finally, with seven or eight simultaneous targets, the probability of cell death peaks at 99\%  \cite{weigelin1,weigelin}. After a tumor cell dies it is removed instantaneously from the lattice. 

\textbf{T cell proliferation.} After releasing the cytotoxic load upon hitting a target, CAR-T cells enter the duplication phase. During this stage, CAR-T cells remain stationary, neither moving nor targeting, as they prepare for duplication. After a 24-hour period, the CAR-T cell divides into two: the original cell and a new CAR-T cell. The original cell retains its remaining lifespan, while the new CAR-T cell is born with 336 hours of lifespan. Activated CAR-T cells have a maximum number of duplications, the so-called Hayflick limit, assumed to be 20 in this work and in line with biological evidence, although this number may depend on the patient’s age and immunological status \cite{Ndifon}. The new CAR-T cell inherits the duplication count from the original cell. For instance, if the original cell has undergone 15 duplications, the count will be 16 in the subsequent time interval for both cells. Upon reaching the maximum number of duplications, the CAR-T cell ceases to divide. We have also studied scenarios with duplication periods different from 24 h, with results listed in the \nameref{S1_Appendix}.

\begin{table}[H]
\caption{\centering Summary of biological parameter values used for the cellular automata.}
\centering
\renewcommand{\arraystretch}{1.5}
\begin{tabular}{|c|c|c|c|}
\hline
\textbf{Parameter} & \textbf{Value} & \textbf{Units} & \textbf{Source}           \\ \hline
$\rho$                              & 1/1000               & hours$^{-1}$                    & \cite{victor}             \\ \hline
$\tau$                              & 336                             & hours                           &  \cite{ghorashian,westera}      \\ \hline
$t_{charge}$                        & 8                               & hours                           & \cite{weigelin1,weigelin} \\ \hline
$n_{dup}$                           & 20                              & dimensionless                   & \cite{weigelin1,weigelin} \\ \hline
$t_{dup}$                           & 24                              & hours                           & \cite{weigelin1,weigelin} \\ \hline
                                    & $0.05$,  if $n=1$                        &                                 &                                            \\
                                    & $0.12$, if $n=2$                           &                                 &                                            \\
$\alpha_n$                          & $0.50$, if $n=3,4$                              & dimensionless                   & \cite{weigelin1,weigelin} \\
                                    & $0.8$, if $n=5,6$                           &                                 &                                            \\
                                    & $0.99$, if $n=7,8$                             &                                 &                                            \\ \hline
\end{tabular}
\label{parametres}
\end{table}

\textbf{Cell motility.} Throughout the simulation, tumor cells remain static at predetermined locations, while CAR-T cells have the ability to move across the lattice. To implement this movement, we considered CAR-T cells to move randomly within their Moore neighborhoods once per time interval, in this case, one hour. Consequently, a CAR-T cell can move to one of the eight adjacent cells as long as they are unoccupied. CAR-T cells with no available space in their neighboorhood, remain in its current location. 

\subsection*{Computational implementation}
The model was implemented in MATLAB version: 9.12 (R2022a). To enhance the reliability of our analyses, we employed a fixed random seed across all simulations. This approach ensured reproducibility, as each rerun of the code produced identical initial cell arrangements. Such consistency facilitated not only the comparison of results but also the robust validation of the model's predictive capabilities. Further details regarding the code and visualization can be found in the Github repository \url{https://github.com/sibordel/CellularAutomataCAR-T.git}.

\section*{Results}

\subsection*{Sparsely distributed tumor cells are more efficiently eliminated by CAR-T cells }
\begin{figure}[H]
    \centering
    \includegraphics[width=1\columnwidth]{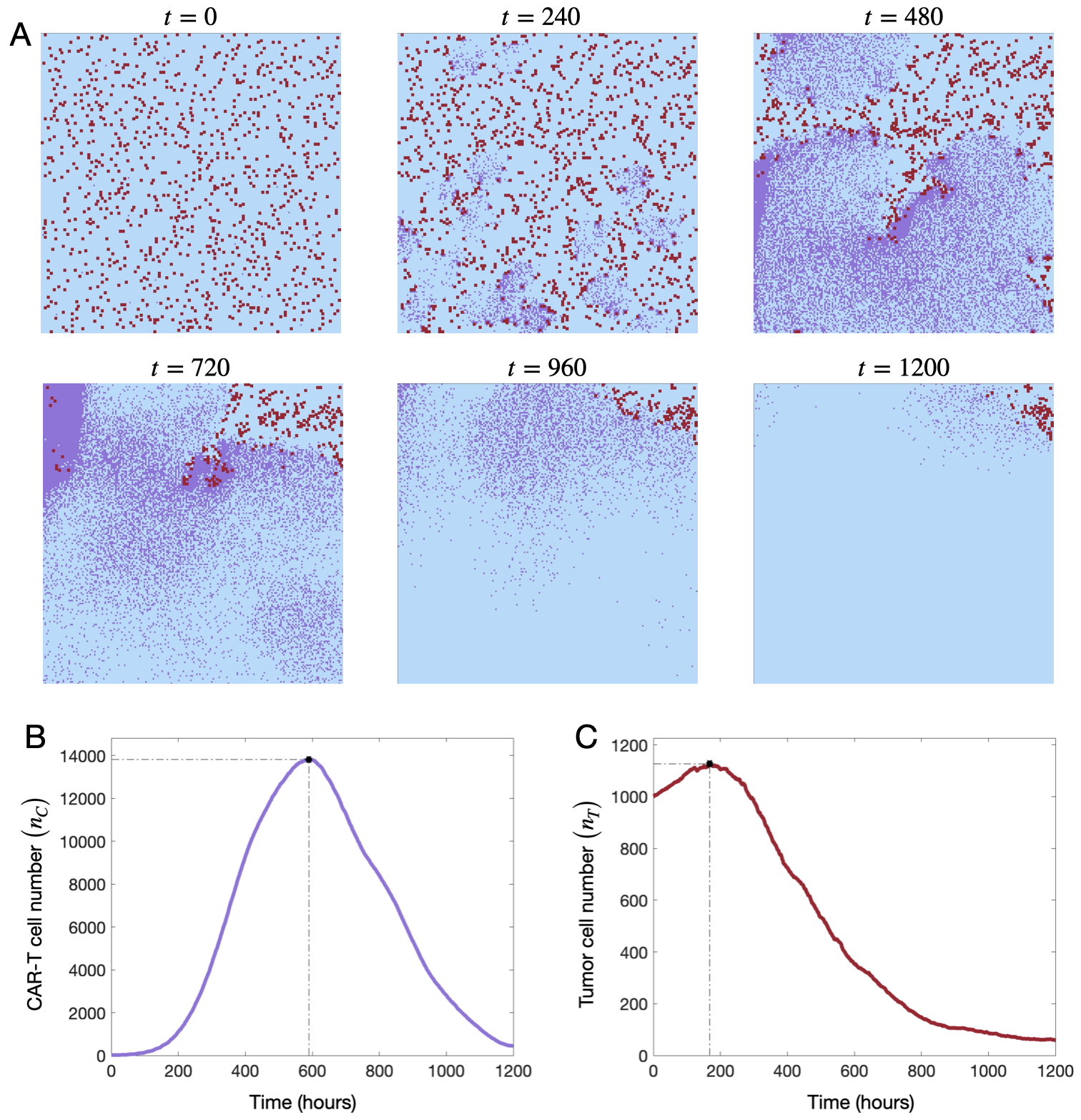}
    \caption{\textbf{Dynamics of CAR-T and tumor cells proliferation with a sparse initial tumor cell distribution.}  \textbf{(A)} Representative images of tumor and CAR-T cell distribution at different timepoints (in hours) during the simulation. \textbf{(B-C)} Expansion kinetics of the CAR-T \textbf{(B)} and tumor cell \textbf{(C)} populations. Starting cell numbers for this simulation were 1000 tumor cells and 20 CAR-T cells. CAR-T cells are shown in purple and tumor cells in red.}
    \label{simulation_sparse}
\end{figure}

To establish the dynamics of CAR-T cell-mediated killing we simulated a sparse distribution of tumor cells randomly located within the computational domain (Fig \ref{simulation_sparse} (A)). In this analysis, the number of CAR-T cells is represented as $n_C$ and the number of tumor cells as $n_T$  with 1000 tumor cells and 20 CAR-T cells initially seeded. This translates to an effector (CAR-T cell) to target (tumor cell) ratio (ET ratio) of 1:100, reflecting a sparsely infiltrated tumor tissue. 

Our simulation shows a strong expansion of the CAR-T cell population after encountering tumor cells (Fig \ref{simulation_sparse} (B)).  The increasing CAR-T cell numbers are accompanied by a decrease in the number of tumor cells over time (Fig \ref{simulation_sparse} (C)), demonstrating the effectiveness of CAR-T cell therapy in targeting and eliminating sparsely distributed tumor cells. The  number of tumor cells peaks at hour 168, just before initiating a robust decline, indicating the timepoint at which CAR-T cells reached a sufficient density to effectively target and eliminate the  tumor cells. This correlated with an effector to target cell ratio (ET ratio) of approximately 1:15.

\begin{figure}[H]
    \centering
    \includegraphics[width=1\columnwidth]{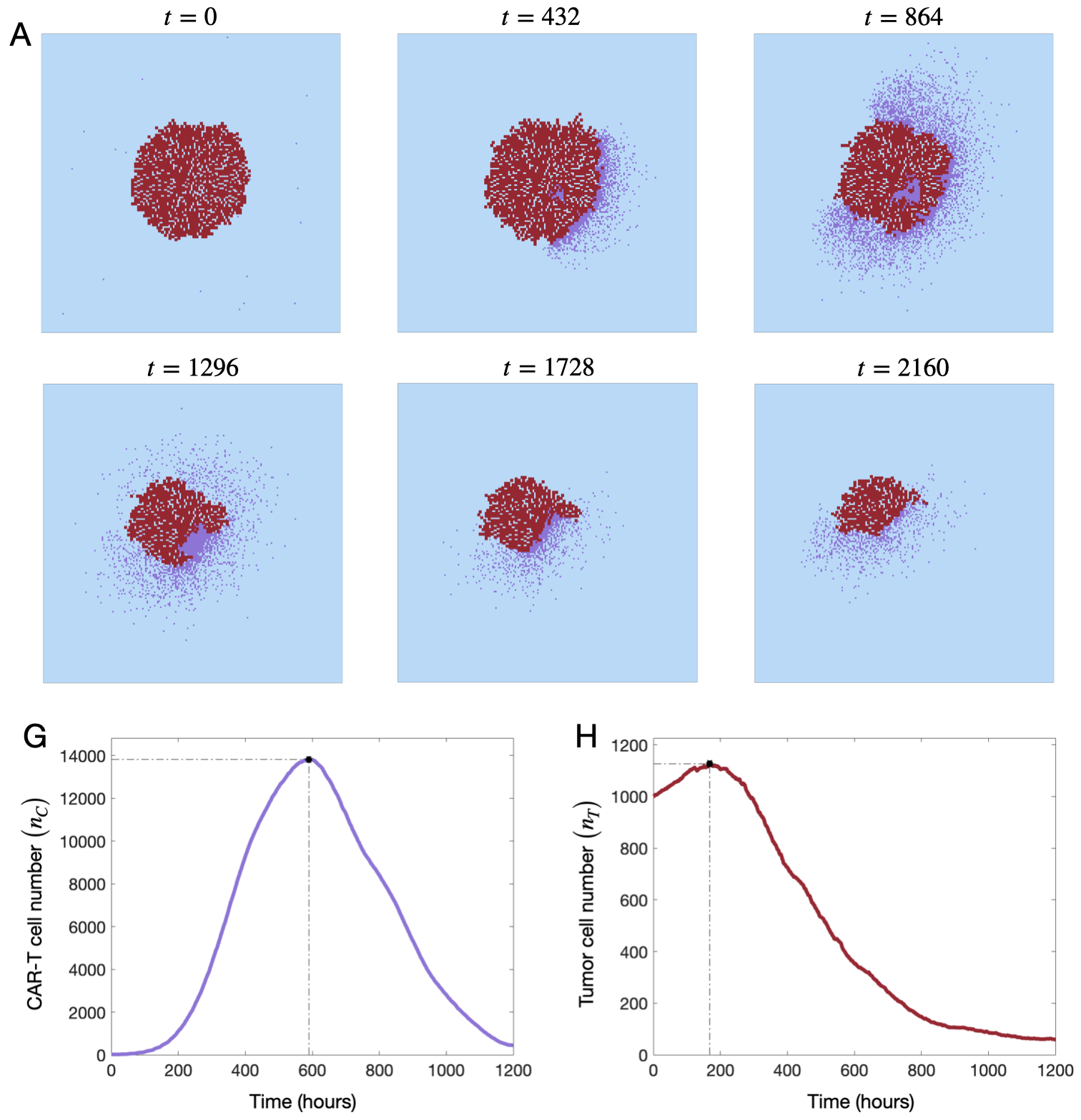}
    \caption{\textbf{Dynamics  of CAR-T and tumor cells with a dense and compact initial tumor cell distribution.} \textbf{(A)} Representative images of tumor and CAR-T cell distribution at different timepoints (in hours) during the simulation. \textbf{(B, C)} Expansion kinetics of the CAR-T \textbf{(B)} and tumor cell \textbf{(C)} populations. Starting cell numbers: 1000 tumor cells and 20 CAR-T cells. CAR-T cells are shown in purple and tumor cells in red.}
    \label{simulation_block}
\end{figure}

To address how the geometry of tumor cell positioning affects CAR-T cell-mediated killing of tumor cells, we proceeded to study a scenario where tumor cells were arranged in a cell-dense, compact solid tumor-like configuration (Fig \ref{simulation_block}(A)). Similarly to the setup with sparse tumor cell distribution, we initially seeded 1000 tumor cells and $20$ CAR-T cells. Tumor cells were arranged in a compact shape located at the center of the automata (Fig \ref{simulation_block}(A)) and the simulations covered a period of 90 days ($2160$ hours).

In this scenario, the dense clustering of tumor cells limited physical contacts between CAR-T cells and their targets to the tumor-tissue interface. Consequently, CAR-T cell expansion was significantly reduced, compared to the sparse tumor cell distribution. Specifically, the maximum CAR-T cell count peaked after 2805 hours, significantly later compared to the simulation with sparse tumor cell distribution (after $100$4 hours). Correspondingly, the peak number of tumor cells was recorded at $t = 1048$ hours, significantly later than in the sparse configuration where tumor cell numbers peaked at hour $t = 168$.

\subsection*{Insensitivity of model results to tumor proliferation rates}

\begin{figure}[H]
    \centering
    \includegraphics[width=1\columnwidth]{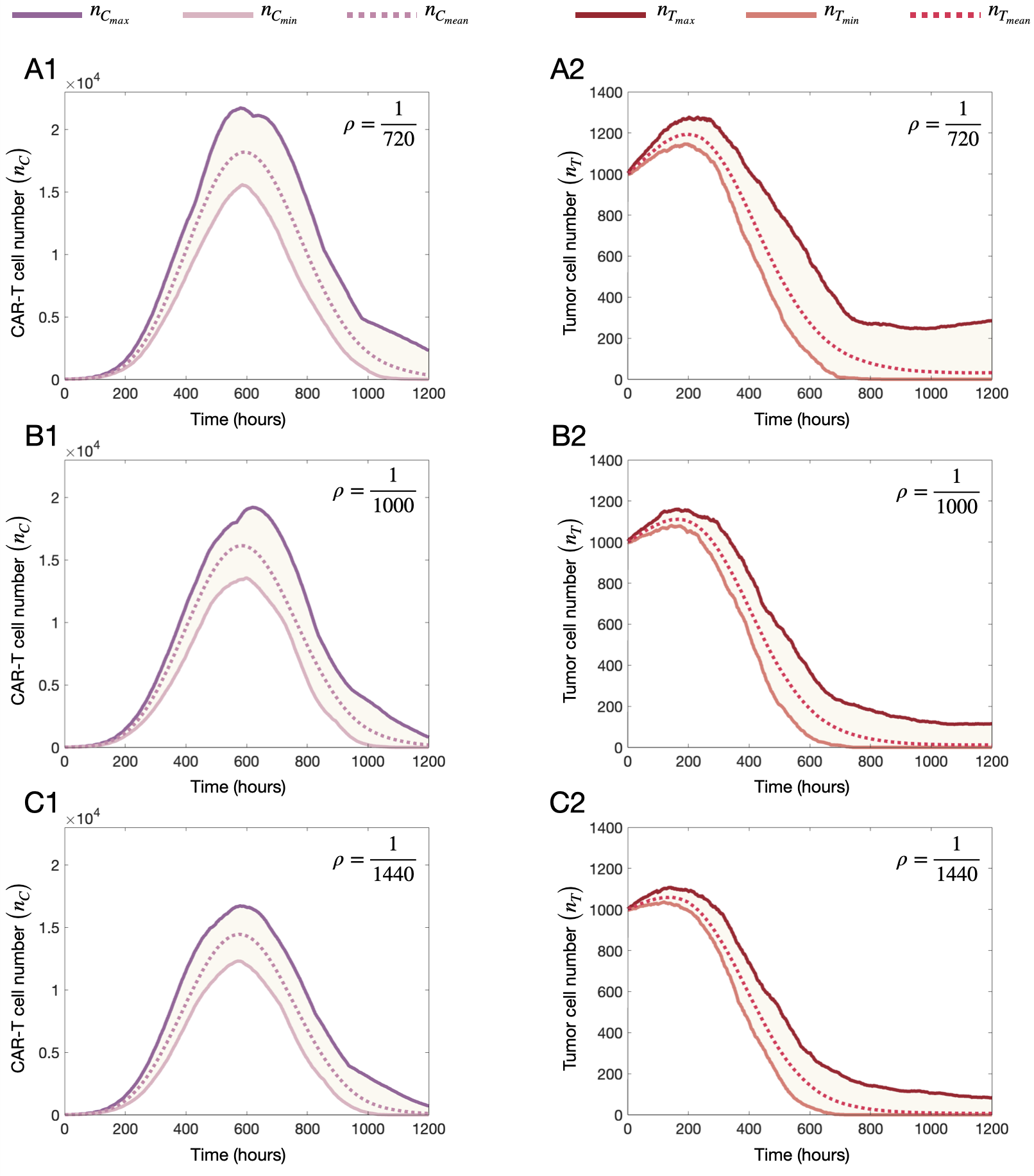}
    \caption{\textbf{Dynamics of sparse tumor cell configurations for different values of $\bm{\rho}$}. All simulations started with 1000 tumor and 20 CAR-T cells. One hundred different random seeds were simulated. The $n_{C_{max}}$ and $n_{T_{max}}$ curves represent the maximum value of CAR-T cells and tumor cells, at each time step of the simulation accross the set of all simulations. $n_{C_{min}}$ and $n_{T_{min}}$ curves represent the minimum values of CAR-T cells and tumor cells at each time step. The $n_{C_{mean}}$ and $n_{T_{mean}}$ curves show the mean of all simulations. Panels display the behavior for different values of $\rho$. (\textbf{A1}, \textbf{A2}) $\rho=1/720$ hours$^{-1}$, (\textbf{B1}, \textbf{B2}) $\rho=1/1000$ hours$^{-1}$, (\textbf{C1}, \textbf{C2})  $\rho=1/1440$ hours$^{-1}$.  CAR-T cell (\textbf{A1, B1, C1}) and tumor cell  (\textbf{A2, B2, C2}) population dynamics are shown. Data from 100 simulations for each set of parameter values.}
    \label{rho_dispersas}
\end{figure}

\begin{figure}[H]
    \centering
    \includegraphics[width=1\columnwidth]{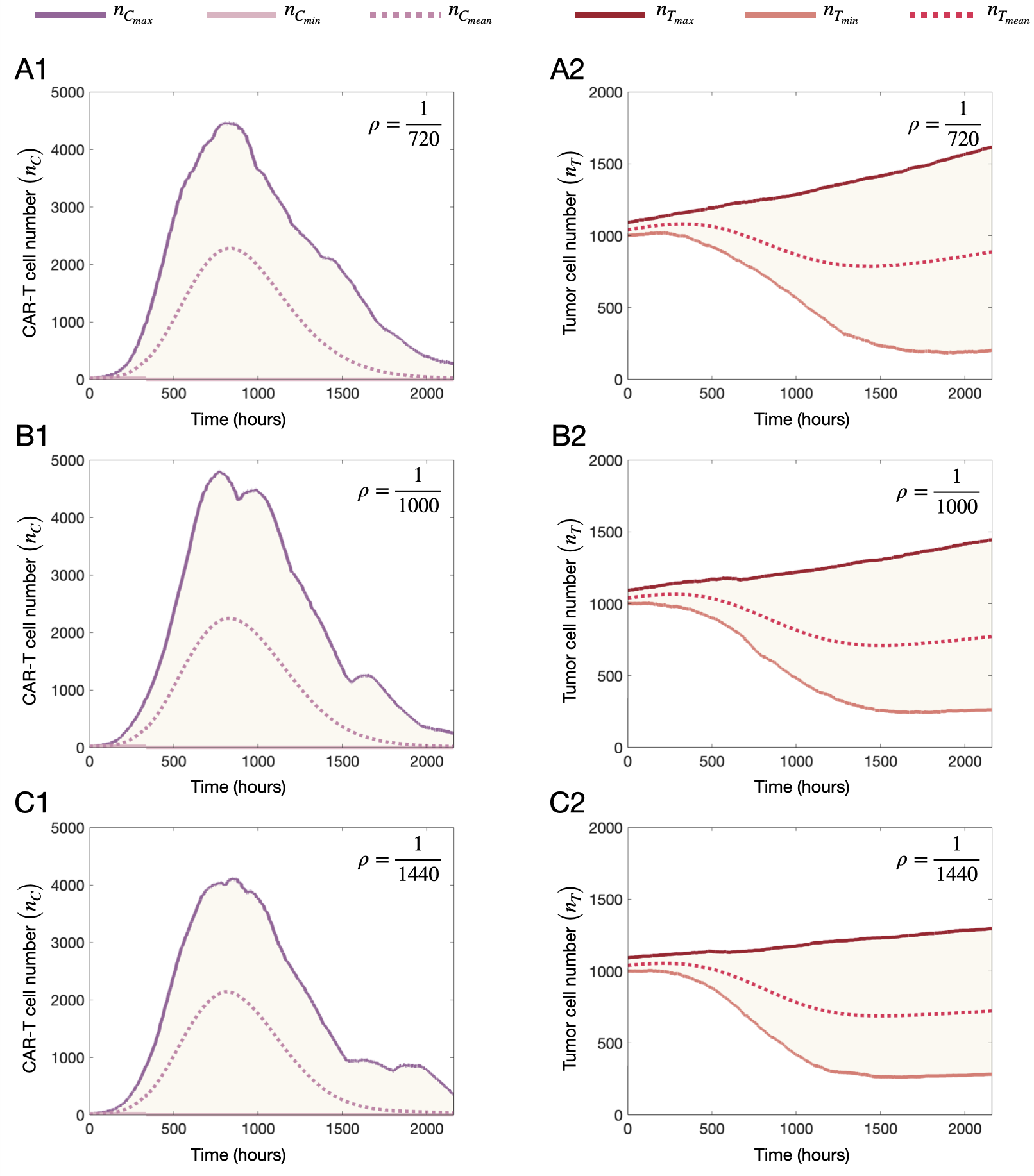}
  \caption{\textbf{Dynamics of compact tumor cell configurations for different values of $\bm{\rho}$}. All simulations started with 1000 tumor and 20 CAR-T cells. One hundred different random seeds were simulated. The $n_{C_{max}}$ and $n_{T_{max}}$ curves represent the maximum value of CAR-T cells and tumor cells, at each time step of the simulation accross the set of all simulations. $n_{C_{min}}$ and $n_{T_{min}}$ curves represent the minimum values of CAR-T cells and tumor cells at each time step. The $n_{C_{mean}}$ and $n_{T_{mean}}$ curves show the mean of all simulations. Panels display the behavior for different values of $\rho$. (\textbf{A1}, \textbf{A2}) $\rho=1/720$ hours$^{-1}$, (\textbf{B1}, \textbf{B2}) $\rho=1/1000$ hours$^{-1}$, (\textbf{C1}, \textbf{C2})  $\rho=1/1440$ hours$^{-1}$. CAR-T cell (\textbf{A1, B1, C1}) and tumor cell  (\textbf{A2, B2, C2}) population dynamics are shown. Data from 100 simulations for each set of parameter values.}
    \label{rho_block}
\end{figure}

In our study, we assumed a constant proliferation rate of tumor cells. Previous research \cite{Ode} explored a range of values for this rate ranging from $1/720$ to $1/1440$ hours$^{-1}$. For most simulations in this paper we chose a rate of $1/1000$ hours$^{-1}$, roughly in the middle of this range. To address, whether this value accurately represented the system or if adjustments would affect our model’s behavior  we explored a broad range of $\rho$ values for tumor cells distributed sparsely (Fig \ref{rho_dispersas}) or as compact tumor mass (Fig \ref{rho_block}).  In both settings, tumor cells were randomly distributed with a fixed number of initially seeded tumor and CAR-T cells. For sparsely distributed and clustered tumor cells,  the evolution of cell populations over time showed comparable trends across all three $\rho$ values. This suggests that the proliferation rate $\rho$ may not have had a significant impact on model dynamics within the investigated parameter range. This relative insensitivity of the model to changes in $\rho$ indicates a robustness in cellular dynamics against fluctuations in this parameter, observed for both sparse and compact tumor cell configurations.

\subsection*{Dependence of treatment efficacy on CAR-T cell homing to the tumor}

CAR-T cell numbers in the tumor depend on CAR-T cell proliferation in the TME and initial CAR-T cell extravasation from the blood vessels. We thus tested how the initial number of CAR-T cells present at the beginning of the simulation affects tumor immune control, simulating varying CAR-T cell extravasation rates in vivo. In these simulations, we fixed the initial tumor cell count to 600 and varied the number of CAR-T cells (10, 20 or 40 CAR-T cells). For both the sparse and compact tumor cell configurations, we conducted 100 simulations using different random seeds for each choice of cell numbers and configuration. 

\begin{figure}[H]
    \centering
    \includegraphics[width=1\columnwidth]{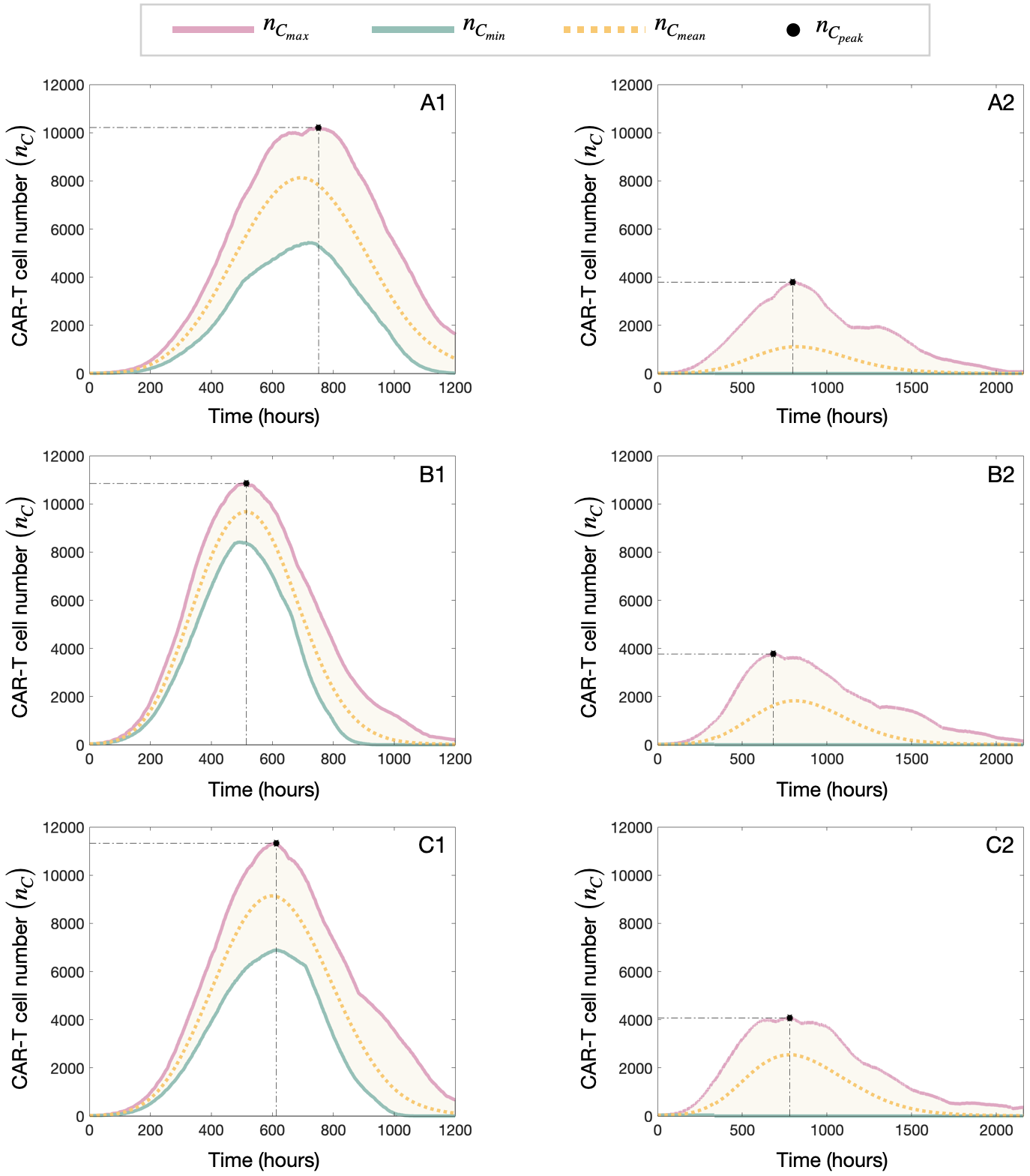}
    \caption{\textbf{Dynamics of CAR-T cell populations for \bm{$n_T=600$} initial tumor cells and varying initial values of CAR-T cells.}  Pink, yellow and green lines represent respectively the maximum, average and minimum values obtained at each time step after conducting $100$ simulations with different initial configurations.  Results are shown for sparse (\textbf{A1}, \textbf{B1}, \textbf{C1}) or compact (\textbf{A2}, \textbf{B2}, \textbf{C2}) initial tumor cell configurations and 10 (\textbf{A1},\textbf{A2}), 20 (\textbf{B1},\textbf{B2}) or 40 (\textbf{C1},\textbf{C2}) initial CAR-T cells. Data from 100 simulations for each set of parameter values.}
    \label{evolution}
\end{figure}

The simulations confirmed the previously observed differences in CAR-T cell expansion and tumor cell elimination in sparse (A1, B1, C1) and compact (A2, B2, C2) tumor cell configurations (Fig \ref{evolution}). The maximum number of CAR-T cells was substantially higher when tumor cells were sparsely distributed, and the peak was reached earlier in the simulation. In simulations with compact tumor configurations, CAR-T cell numbers consistently remained close to zero, corresponding to a large fraction of simulations with minimal CAR-T cell expansion (Fig \ref{evolution}, A2, B2, C2). However, comparing the maximum CAR-T cell numbers and the timepoints the peak was reached showed no significant differences between simulations starting with different CAR-T cells numbers. This indicates that the initial number of CAR-T cells has less impact on the maximum expansion compared to the CAR-T cell-intrinsic proliferation rate.

\subsection*{Initial tumor load determines peak CAR-T expansion and initial number of cells injected influences time to peak}

\begin{figure}[H]
	\centering
	\includegraphics[width=1\columnwidth]{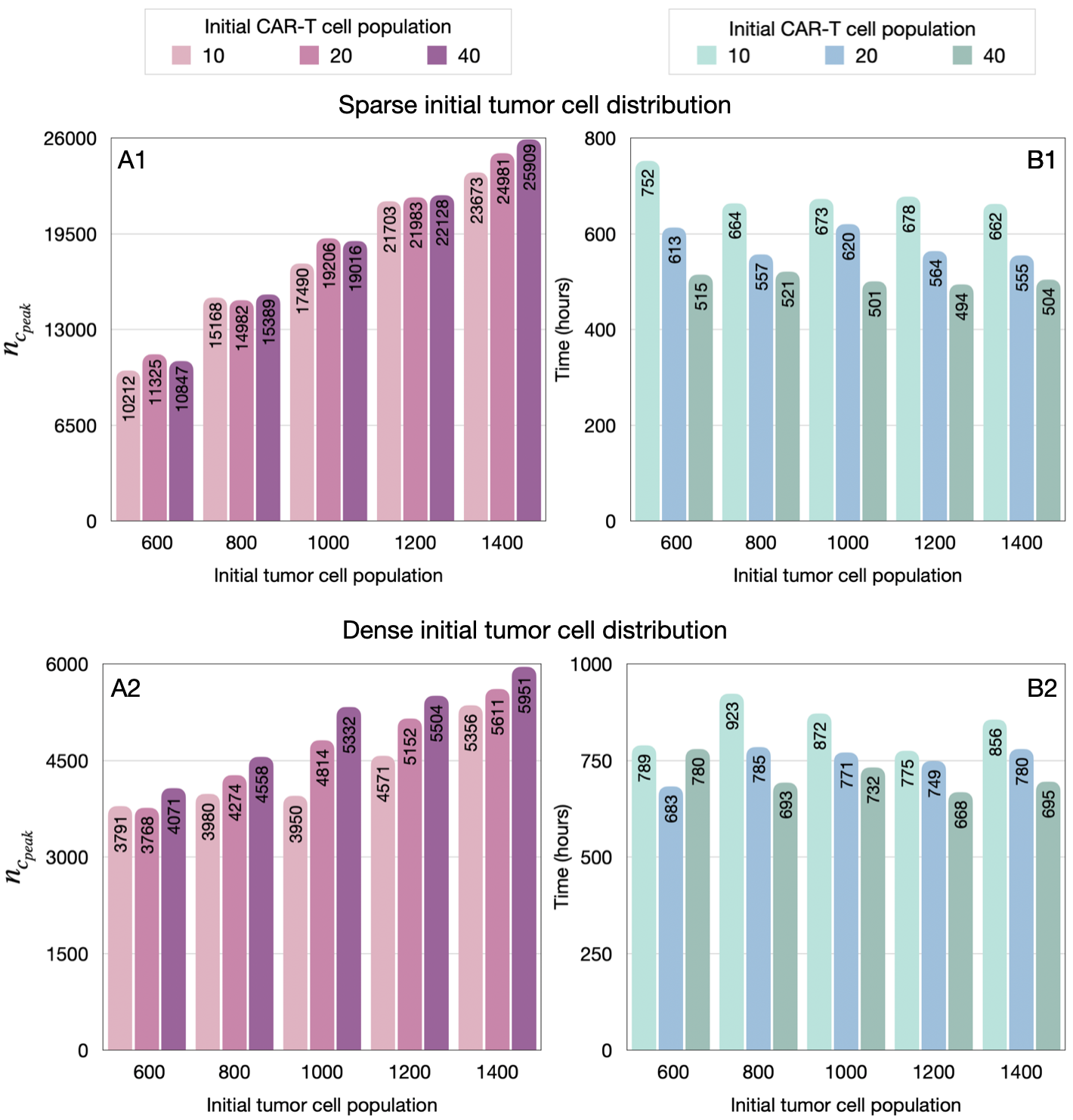}
	\caption{\textbf{Peak number and time of peak expansion of CAR-T cells in sparse and dense tumor cell configurations.} \textbf{(A1, A2)} Maximum number of CAR-T cells obtained for different initial values of CAR-T (10, 20, 40) and tumor cells (600, 800, 1000, 1200, 1400). \textbf{(B1, B2)} Time to peak CAR-T cell expansion for the numbers of CAR-T and tumor cells.}
	\label{barras}
\end{figure}

To quantify the relation between CAR-T cells initially present at the start of the simulation and the impact of tumor cell numbers on the timepoint of maximum CAR-T cell expansion, we plotted the maximum count of CAR-T cells  (Fig \ref{barras}(A1, A2)) and the time when it occurred (Fig \ref{barras}(B1, B2))  in sparsely and clustered tumor cell distributions with increasing numbers of initial tumor cells.

%\textbf{UNIFICAR LAS DOS GRÁFICAS EN LA MISMA Y PONER EL PIE DE FOTO QUE ESTÁ AHORA EN LA FIGURA 7}

%\begin{figure}[H]
%	\centering
%	\includegraphics[width=1\columnwidth]{barras_block.png}
%	\caption{\textbf{Peak levels and times of peak expansion of CAR-T cells in compact tumor cell configurations.} The figure illustrates simulations depicting peak levels and timing for different initial conditions of CAR-T cells and solid tumor cells. \textbf{A.} Maximum peaks of CAR-T cells obtained for different initial values of CAR-T (10,20,40) and tumor cells (600,800,1000,1200,1400). \textbf{B.} Time to peak CAR-T cell expansion for the numbers of CAR-T and tumor cells.}
%	\label{barras_block}
%\end{figure}

For sparsely distributed tumor cells, we found that the maximum number of CAR-T cells increased depending on the number of tumor cells seeded at the beginning of the simulation (Fig \ref{barras}(A1)). While the peak number of CAR-T cells did not depend on the number of CAR-T cells seeded initially, the timing of its occurrence did (Fig \ref{barras}(B1)). We consistently observed an earlier occurrence of the maximum CAR-T cell expansion with increasing initial CAR-T cell numbers. The time required to reach the maximum CAR-T cell expansion was independent on the initial number of tumor cells in the range studied (Fig \ref{barras}(B1)). Consistently, in simulations with clustered tumor cells, the initial number of tumor cells influenced the maximum expansion of CAR-T cells (Fig \ref{barras}(A2)) and the initial number of CAR-T cells modulated the timepoint when the peak expansion was reached (Fig \ref{barras}(B2)), but both effects were weaker than in simulations with sparsely distributed tumor cells. 

In summary, the initial number of tumor cells influenced the maximum expansion of CAR-T cells but not the time until the peak expansion was reached. In contrast, the initial number of CAR-T cells did not influence the maximum expansion of CAR-T cells but accelerated the time until the peak expansion was reached.

\subsection*{Initial states determine therapy success or failure}

\begin{figure}[H]
    \centering
    \includegraphics[width=0.95\columnwidth]{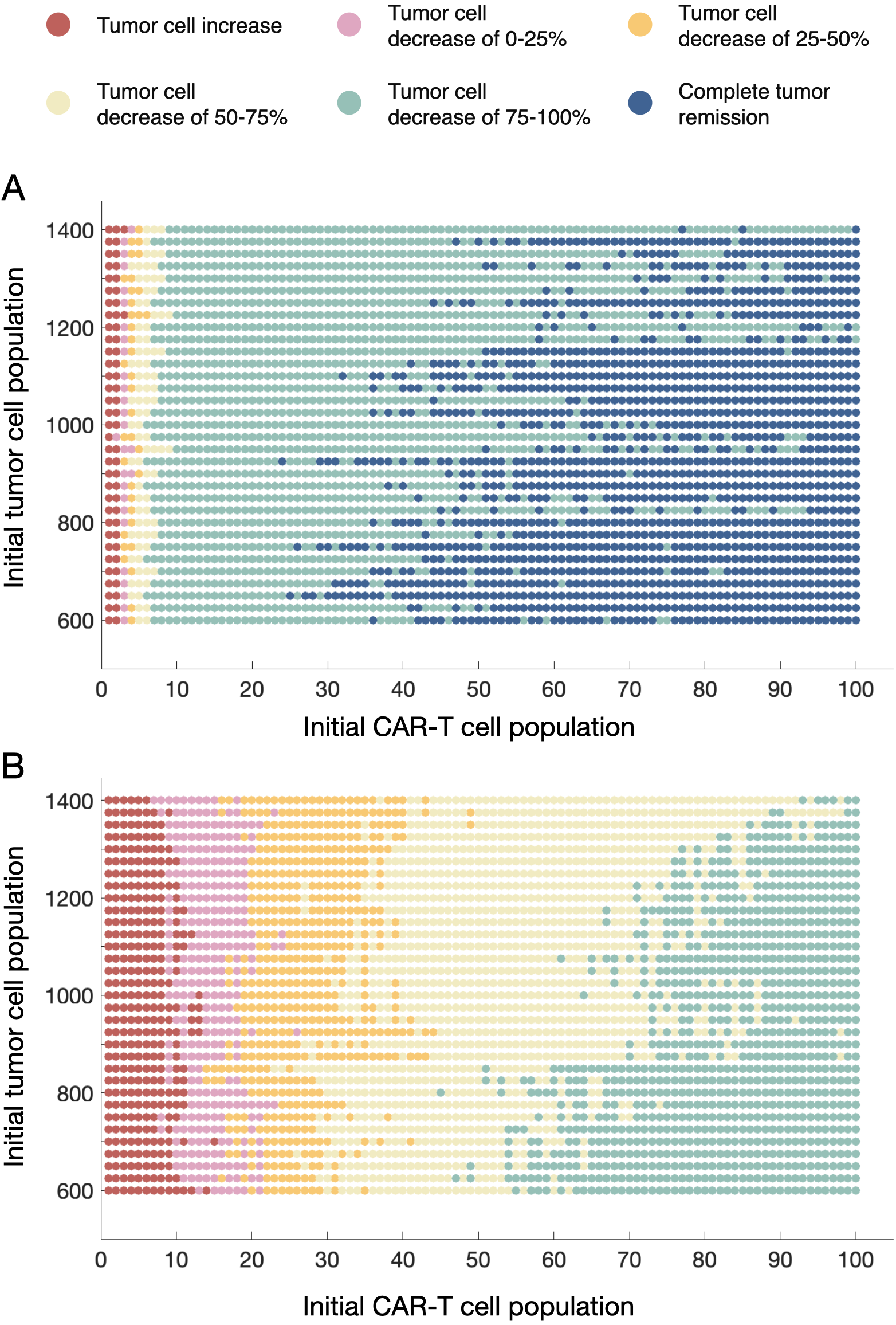}
    \caption{\textbf{Impact of the initial number of tumor cells, CAR-T cells, and tumor geometry on treatment outcome.} Sparse \textbf{(A)} and compact \textbf{(B)} tumor geometries were explored. For each value of the tumor cell number in the range 600 to 1400 cells (Y axis) and CAR-T cell number in the range 1 to 100 (X axis) each point indicates the outcome of 20 simulations with different initial random seeds. Red indicate growth of the mean of the simulations in comparison with the initial number of tumor cells, pink denotes $<25$\% reduction, orange represents 25-50\% reduction, yellow corresponds to 50-75\% reduction, green indicates $>75$\% reduction, and blue denotes complete remission in all of the 20 simulations.}
    \label{survival}
\end{figure}

Finally, to comprehensively evaluate tumor response to CAR-T cell treatment at the end of the simulation for sparsely and clustered tumor cell distribution and the initial quantities of tumor and CAR-T cells, we applied extensive simulations (Fig \ref{survival} (A, B)). For each configuration and pair of values of the initial tumor and CAR-T cell numbers we run 20 simulations and classified the response for each choice of the initial number in either complete response (elimination of all cancer cells in all of the 20 simulations), no response (mean growth of tumor cells) and different categories depending on whether the mean tumor cell number reduction at the end of the simulation was between 0\% - 25\%, 25\% - 50\%, 50\% - 75\%, 75\% - 100\%. A total of 33 different values of the initial tumor load and 100 values of the initial number of CAR-T cells where studied leading to a total of 66000 simulations. We observed a full range of therapy response, ranging from complete tumor remission to immune evasion indicated by continued tumor growth (Fig \ref{survival} (A, B)).

The count of CAR-T cells increases incrementally from 1 to 100, while the number of tumor cells varies in increments of 25, ranging from 600 to 1400. Consequently, each graph comprises a total of 330 data points.

Complete tumor remission was observed with higher frequency in simulations where tumor cells were sparsely distributed, while complete remission was never observed in clustered tumor cells (Fig \ref{survival} (A, B)). Furthermore, in the case of sparse tumor configurations, treatment appears to be largely effective, as indicated by the scarcity of simulations resulting in overall tumor growth or response rates below 50\% response rate (Fig \ref{survival} (A)). Conversely, in simulations starting with already clustered tumor cells, treatment efficacy is diminished, with minimal tumor reduction observed in the majority of cases (Fig \ref{survival} (B)). 
In both tumor geometries, efficacy was dependent on CAR-T cell numbers at the beginning of the simulation. Even sparsely distributed tumor cells could be controlled when less than 25 CAR-T cells were present at the beginning of the simulation, representing a starting ET ratio ranging from 1:24 - 1:56, depending on the number of tumor cells present. In contrast, 94\% of the tumors were eradicated when the starting ET ratio ranged from 1:6 - 1:14. In compact tumors, no tumor decreased more than 50\% at ET ratios ranging from 1:14 - 1:31 or lower while tumor response rates ranged between 75\% - 100\% at ET ratios of 1:6 - 1:14. Interestingly there was a sustained trend in reduction of CAR-T cell effectiveness in compact tumors with increasing numbers of tumor cells which was not observed for sparsely distributed tumor cells (Fig \ref{survival} (A, B)). The perimeter to surface area decreases with increasing radius and thus indicates tumor size as a major determinant of treatment failure when CAR-T cells infiltrate the tumor from the tumor-tissue interfaces.

\section*{Discussion}

%Decir que mucho no penetran en el tumor y por eso vale 
% Morfológicamente irregulares podrían ser mejores respondedores

The interest in mathematical descriptions of immunotherapy treatments for cancer, specifically targeting CAR-T cells, has been steadily growing. This trend reflects a recognized need to enhance our understanding of the intricate dynamics involved in immunotherapeutic approaches that involve in-patient expansion of the drug, exhaustion, phenotype changes between the different T lympocytes injected and their competition with the target tumor cells and the off-tumor on target effects as well as treament toxicity. Mathematical modeling provides a valuable tool for dissecting the complex interactions between effector cells and target cells, offering insights into treatment efficacy and suggesting potential optimization strategies. This study contributes to the ongoing efforts to leverage mathematical frameworks for advancing cancer immunotherapy research.

Most mathematical studies available up to now have focused on the global interactions between the effector and target cells by using compartimental differential equation models to describe CAR-T cell immunotherapy \cite{alt, barros, kim, Liu, salvi, jesus, Ode, Mo, barros1, ivana, victor, roe, mahlbacher}. However, the interactions between T-lymphocytes and cancer cells require a direct physical interaction between both types of cells. This is why in this study we considered spatial aspects of the interactions using a cellular automata. This methodology enabled us to elucidate the nuanced cellular-level dynamics, yielding valuable insights into the mechanisms governing CAR-T cell therapy. Grounded in cellular automata, our modeling approach facilitated a comprehensive exploration of the spatial dimension in cellular dynamics. Unlike conventional models reliant on differential equations, do not account for spatial interactions, our method provided a more realistic depiction of cell distribution and organization within tumor tissue.

Consequently, here we intended to provide a complementary perspective in precisely evaluating the impact of tumor geometry on the effectiveness of CAR-T cell immunotherapy. Our focus was on studying how different tumor geometries could influence the effectiveness of CAR-T cell therapy, thus providing some sort of geometrical immune suppression mechanism that could add to others known to be present in solid tumors.

Our overarching objective was to enhance our understanding of the dynamics and outcomes associated with this treatment approach by investigating diverse scenarios encompassing varying initial counts of tumor cells and CAR-T cells. An important finding emerging from this research highlights the significant impact of tumor geometry on treatment effectiveness, revealing notable differences in responses between tumors characterized by compact cellular arrangements and those with sparse cells configurations. This disparity can be directly attributed to the increased frequency of interactions between tumor and CAR-T cells within sparse tumors. By fostering a higher initial contact density, sparse tumors facilitate greater opportunities for cell duplication, resulting in a more robust expansion of the CAR-T cell population compared to compact tumors. This observation not only aligns with clinical evidence showing better outcomes of CAR-T cell therapy in hematologic malignancies like leukemias, lymphomas, and myelomas \cite{HM, HM1} and highlights another effect that could limit the efficacy of these treatments on solid tumors. Despite ongoing advancements, effectively translating these therapeutic benefits to solid tumors remains a formidable challenge \cite{ST, ST1}.

%The duplication of CAR-T cells upon encountering tumor cells is based on our assertion that antigen-specific stimulation is the primary driver of CAR-T cell proliferation. CAR-T cells are engineered to express chimeric antigen receptors (CARs) targeting specific antigens on cancer cells. Upon antigen binding, CARs initiate intracellular signaling pathways within CAR-T cells, leading to their activation and subsequent proliferation. While cytokines such as interleukin-2 (IL-2), interleukin-7 (IL-7), and interleukin-15 (IL-15) can support CAR-T cell function and proliferation, antigen-specific stimulation provides a more direct and specific signal, driving CAR-T cell activation and proliferation. This highlights the critical role of antigen recognition in CAR-T cell therapy and offers valuable insights for refining CAR-T cell-based immunotherapy in cancer treatment.

Our observations highlight an interesting aspect regarding the initial population of CAR-T cells and its temporal impact on treatment dynamics. We noted that the initial number of CAR-T cells notably influences the time required to reach the peak CAR-T cell count in the system. This implies that the temporal dynamics of treatment are influenced by the number of injected cells. More precisely, we found that a larger initial population of CAR-T cells correlated with a shorter time to reach the peak CAR-T cell count. Even more relevant is the fact that there was a sustained trend in reduction of effectiveness of the treatment in compact tumors as the number of cancer cells increased that was not as evident in sparse tumors. It is clear that the perimeter to surface area decreases with the radius and thus lesion size would be a major determinant of treatment failure in compact solid tumors. The combination of CAR-T cells with treatments that make tumors more accessible to T-cells could be crucial for success. For example, radiopharmaceutical therapies (RPTs) could use the same surface molecule / antigen as the CAR product. Due to their different mechanisms of action, CAR-T treatments could be more effective after RPTs, as the increased accessibility to internal tumor areas created by RPTs would render solid tumors more susceptible to CAR-T cells. High-risk localized prostate cancer has a high recurrence rate due to failed local control and micrometastatic disease. Lutetium PSMA, a radioligand targeting PSMA-expressing prostate cancer cells, has shown efficacy in metastatic castration-resistant prostate cancer. Combining lutetium PSMA with systemic immunotherapies could enhance the anti-tumor T-cell response, potentially improving long-term immunity and targeting occult micrometastases \cite{eapen}.

However, it is essential to acknowledge the limitations inherent in our study. The cellular automata model we utilized represents only a small fraction of the tissue, specifically 4 mm$^2$. It could be relevant to extend the model to a larger scale, encompassing broader tissue areas, including vessels and other elements of the tumor microenvironment. Transitioning to a larger scale would permit a more thorough assessment of how cellular processes and tissue interactions manifest at a macroscopic level. Furthermore, given that our current cellular automata operates within a two-dimensional domain, consideration must be given to advancing to a three-dimensional model. This potential evolution would enable a more accurate depiction of the spatial complexity of tumors and cellular interactions, ultimately enhancing the clinical relevance and precision of our modeling approach. These deliberations underscore the importance of ongoing exploration and refinement of our modeling methodologies to achieve a more holistic and precise representation of biological systems in the context of CAR-T cell therapy.

In our modeling approach, we have simplified our representation by assuming that tumor cells do not experience changes over time. However, this oversimplification fails to capture the dynamic complexity of tumors, where genetic mutations are known to occur, although probably on longer timescales. Such mutations can lead to the emergence of therapy-resistant cell phenotypes, posing challenges to treatment efficacy. This phenomenon, commonly referred to as tumor escape, underscores the importance of considering the dynamic evolution of tumor cells in therapeutic strategies. Incorporating genetic variability and tumor evolution into simulation models could enhance our understanding of disease dynamics and facilitate the development of more effective treatment approaches capable of addressing tumor escape in cancer therapy.

\section*{Conclusion}

In summary, our study highlights the relevance of considering spatial aspects when describing CAR-T cell immunotherapy mathematically. We have found a key role of the tumor geometry indicating a mechanism of ‘geometrical’ immune suppression based on the accessibility of tumor cells to infiltrating immune cells. We have also highlighted the relevant effect of tumor size and number of CAR-T cells homing to the tumor for the control of compact tumors, but not on sparse tumors that are easy to access for the curative effector cells. Moving forward, continued refinement and validation of cellular automata models will be essential to fully harness their potential and facilitate the development of more effective CAR-T cell therapies.  

\section*{Supporting information}
\paragraph*{S1 Supplementary Information.}
\label{S1_Appendix}
{Time variation in CAR-T cell proliferation.}

\section*{Data availability}

All relevant data are within the manuscript. The cellular automata code is available at \url{https://github.com/sibordel/CellularAutomataCAR-T.git}.

\section*{Author contributions}

\textbf{Conceptualization}: Silvia Bordel-Vozmediano, Víctor M. Pérez-García.
%\textbf{Data Curation}: 
\textbf{Formal Analysis}: Silvia Bordel-Vozmediano.
\textbf{Funding Acquisition}: Víctor M. Pérez-García.
\textbf{Investigation}: Silvia Bordel-Vozmediano, Soukaina Sabir, Lucía Benito-Barca, Bettina Weigelin, Víctor M. Pérez-García.
\textbf{Methodology}: Silvia Bordel-Vozmediano, Soukaina Sabir, Víctor M. Pérez-García.
\textbf{Project Administration}: Víctor M. Pérez-García.
%\textbf{Resources}:
\textbf{Software}: Silvia Bordel-Vozmediano, Lucía Benito-Barca.
\textbf{Supervision}: Víctor M. Pérez-García.
%\textbf{Validation}:
\textbf{Visualization}:	Silvia Bordel-Vozmediano
\textbf{Writing – Original Draft}: Silvia Bordel-Vozmediano, Soukaina Sabir.
\textbf{Writing – Review \& Editing}: Silvia Bordel-Vozmediano, Soukaina Sabir, Lucía Benito-Barca, Bettina Weigelin, Víctor M. Pérez-García.

\section*{Acknowledgements}

This work has been partially supported by project PID2022-142341OB-I00, funded by Ministerio de Ciencia e Innovación/Agencia Estatal de Investigación. Spain (doi:10.13039/501100011033) and European Regional Development Fund (ERDF A way of making Europe);  grant SBPLY/21/180501/000145 (Junta de Comunidades de Castilla-La Mancha, Spain) and grant 2022-GRIN-34405 funded by University of Castilla-La Mancha/FEDER (Applied Science Projects within the UCLM research programme).

\nolinenumbers


\begin{thebibliography}{99}

\bibitem{Waldman} Waldman AD, Fritz JM, Lenardo MJ. A guide to cancer immunotherapy: from T cell basic science to clinical practice. Nat Rev Immunol 2020; 20:651–668. http://dx.doi.org/10.1038/s41577-020-0306-5

\bibitem{June} June CH, O'Connor RS, Kawalekar OU, Ghassemi S, Milone MC. CAR-T cell immunotherapy for human cancer. Science 2018; 359(6382):1361-1365. http://dx.doi.org/10.1126/science.aar6711

\bibitem{Sterner} Sterner RC, Sterner RM. CAR-T cell therapy: current limitations and potential strategies. Blood Cancer J 2021; 11:69. http://dx.doi.org/10.1038/s41408-021-00459-7


\bibitem{Zhang} Zhang X, Zhu L, Zhang H, Chen S, Xiao Y. CAR-T Cell Therapy in Hematological Malignancies: Current Opportunities and Challenges. Front Immunol. 2022; 13:927153. http://dx.doi.org/10.3389/fimmu.2022.927153.

\bibitem{Lu} Lu J, Jiang G. The journey of CAR-T therapy in hematological malignancies. Mol Cancer 2022; 21:194. http://dx.doi.org/10.1186/s12943-022-01663-0


\bibitem{Uslu} Uslu U, June CH. Beyond the blood: expanding CAR T cell therapy to solid tumors. Nat Biotechnol 2024; 1–10. http://dx.doi.org/10.1038/s41587-024-00123-4
 	 

\bibitem{Albelda} Albelda SM. CAR-T cell therapy for patients with solid tumours: key lessons to learn and unlearn. Nat Rev Clin Oncol. 2024; 21(1):47-66. http://dx.doi.org/10.1038/s41571-023-00832-4.

\bibitem{Guzman} Guzman G, Reed MR, Bielamowicz K, Koss B, Rodriguez A. CAR-T therapies in solid tumors: opportunities and challenges. Curr Oncol Rep 2023; 25(5):479–489. http://dx.doi.org/10.1007/s11912-

\bibitem{Schmidts} Schmidts A, Maus MV. Making CAR T cells a solid option for solid tumors. Front Immunol 2018; 9:2593. http://dx.doi.org/10.3389/fimmu.2018.02593

\bibitem{barros} Barros LRC, Paixão EA, Valli AMP, Naozuka GT, Fassoni AC, Almeida RC. CARTmath—a mathematical model of CAR-T immunotherapy in preclinical studies of hematological cancers. Cancers 2021;13(12):2941. https://doi.org/10.3390/cancers13122941

\bibitem{alt} Kimmel GJ, Locke FL, Altrock PM. The roles of T cell competition and stochastic extinction events in chimeric antigen receptor T cell therapy. Proc R Soc B 2021; 288(1947). http://dx.doi.org/10.1098/rspb.2021.0229


\bibitem{Ode} León-Triana O, Sabir S, Calvo GF, et al. CAR-T cell therapy in B-cell acute lymphoblastic leukaemia: Insights from mathematical models. Comm Nonlinear Sci Numer Simul. 2021;94:105570.  http://dx.doi.org/10.1016/j.cnsns.2020.105570

\bibitem{Liu}  Liu L, Ma C, Zhang Z, et al. Computational model of CAR-T-cell immunotherapy dissects and predicts leukemia patient responses at remission, resistance, and relapse. J Immunother Cancer. 2022;10(12):e005360.  http://dx.doi.org/10.1136/jitc-2022-005360

\bibitem{salvi} Martínez-Rubio Á, Chulián S, Blázquez Goñi C, et al. A Mathematical Description of the Bone Marrow Dynamics during CAR-T-Cell Therapy in B-Cell Childhood Acute Lymphoblastic Leukemia. Int J Mol Sci. 2021;22(12):6371. http://dx.doi.org/10.3390/ijms22126371

\bibitem{Mo} Mostolizadeh R, Afsharnezhad Z, Marciniak-Czochra A. Mathematical model of Chimeric Anti-gene Receptor (CAR) T cell therapy with presence of cytokine. Numerical Algebra, Control \&amp; Optimization. 2018;8(1):63–80. http://dx.doi.org/10.3934/naco.2018004

\bibitem {ivana} Owens K, Bozic I. Modeling CAR-T-Cell Therapy with Patient Preconditioning. Bull Math Biol. 2021;83(5). http://dx.doi.org/10.1007/s11538-021-00869-5

\bibitem {victor}  Pérez-García VM, León-Triana O, Rosa M, Pérez-Martínez A. CAR-T cells for T-cell leukemias: Insights from mathematical models. Comm Nonlinear Sci Numer Simul. 2021;96:105684. http://dx.doi.org/10.1016/j.cnsns.2020.105684

\bibitem{roe} Roesch K, Hasenclever D, Scholz M. Modelling Lymphoma Therapy and Outcome. Bull Math Biol. 2013;76(2):401–430. http://dx.doi.org/10.1007/s11538-013-9925-3

\bibitem{juan} Bodnar M, Foryś U, Piotrowska MJ, Bodzioch M, Romero-Rosales JA, Belmonte-Beitia J. On the analysis of a mathematical model of CAR-T cell therapy for glioblastoma: Insights from a mathematical model. Int J Appl Math Comp Sci. 2023. https://doi.org/10.34768/amcs-2023-0027

\bibitem{brain} León-Triana O, Pérez-Martínez A, Ramírez-Orellana M, Pérez-García VM. Dual-Target CAR-Ts with On- and Off-Tumour Activity May Override Immune Suppression in Solid Cancers: A Mathematical Proof of Concept. Cancers. 2021;13(4):703.  http://dx.doi.org/10.3390/cancers13040703

\bibitem{Russel} Li R, Sahoo P, Wang D, et al. Modeling interaction of Glioma cells and CAR-T-cells considering multiple CAR-T-cells bindings. ImmunoInformatics. 2023;9:100022. http://dx.doi.org/10.1016/j.immuno.2023.100022

\bibitem{Sahoo} Sahoo P, Yang X, Abler D, et al. Mathematical deconvolution of CAR T-cell proliferation and exhaustion from real-time killing assay data. J R Soc Interface. 2020;17(163):20190734. https://doi.org/10.1098/rsif.2019.0734

\bibitem{Owens} Owens K, Rahman A, Bozic I. Spatiotemporal dynamics of tumor - CAR T-cell interaction following local administration in solid cancers. 2024. https://doi.org/10.1101/2024.08.29.610392

\bibitem{CA1} Burks A W. Essays on Cellular Automata. University of Illinois Press. 1968.

\bibitem{CA} Edmundson HP. Theory of self-reproducing automata. Information Storage and Retrieval. 1969; 5(3):151. http://dx.doi.org/10.1016/0020-0271(69)90026-6

\bibitem{ger} Gerlee P, Anderson ARA. An evolutionary hybrid cellular automaton model of solid tumour growth. J Theor Biol. 2007; 246(4):583–603.  http://dx.doi.org/10.1016/j.jtbi.2007.01.027

\bibitem{mallet} Kazmi N, Hossain MA, Phillips RM. A Hybrid Cellular Automaton Model of Solid Tumor Growth and Bioreductive Drug Transport. IEEE/ACM Trans. Comput. Biol. and Bioinf. 2012;9(6):1595–1606.  http://dx.doi.org/10.1109/TCBB.2012.118

\bibitem{pata} Byrne HM. Dissecting cancer through mathematics: from the cell to the animal model. Nat Rev Cancer. 2010;10(3):221–230. http://dx.doi.org/10.1038/nrc2808

\bibitem{mallet1}Mallet DG, De Pillis LG. A cellular automata model of tumor–immune system interactions. J Theor Biol. 2006;239(3):334–350.  http://dx.doi.org/10.1016/j.jtbi.2005.08.002

\bibitem{zouhri} Zouhri S, Saadi S, Rachik M. Simulation of Tumor Response to Immunotherapy Using a Hybrid Cellular Automata Model. Int. J. Appl. Comput. Math. 2016;3(2):1077–1101.  http://dx.doi.org/10.1007/s40819-016-0163-x

\bibitem{weigelin1}Weigelin B, Boer ATh den, Wagena E, et al. Cytotoxic T cells are able to efficiently eliminate cancer cells by additive cytotoxicity. Nat. Commun. 2021;12(1). http://dx.doi.org/10.1038/s41467-021-25282-3

\bibitem{weigelin}Weigelin B, Friedl P. T cell-mediated additive cytotoxicity – death by multiple bullets. Trends in Cancer. 2022;8(12):980–987. http://dx.doi.org/10.1016/j.trecan.2022.07.007

\bibitem{Ndifon} Ndifon W, Dushoff J. The Hayflick limit may determine the effective clonal diversity of naive T cells. J Immunol 2016; 196(12):4999–5004. http://dx.doi.org/10.4049/jimmunol.1502343

\bibitem{westera} Westera , Drylewicz J, den Braber I, Mugwagwa T, van der Maas I, Kwast L, Volman T, van de Weg-Schrijver EH, Bartha I, Spierenburg G, Gaiser K, Ackermans MT, Asquith B, de Boer RJ, Tesselaar K, Borghans JA. Closing the gap between T-cell life span estimates from stable isotope-labeling studies in mice and humans. Blood. 2013 Sep 26;122(13):2205-12. http://dx.doi.org/10.1182/blood-2013-03-488411

\bibitem{ghorashian} Ghorashian S, Kramer AM, Onuoha S. Enhanced CAR T cell expansion and prolonged persistence in pediatric patients with ALL treated with a low-affinity CD19 CAR. Nat Med 2019; 25:1408–1414. http://dx.doi.org/10.1038/s41591-019-0549-5

\bibitem{jesus} de Jesus Rodrigues B, Barros LRC, Almeida RC. Three-compartment model of CAR-T-cell immunotherapy. Cold Spring Harb Lab 2019. http://dx.doi.org/10.1101/779793

\bibitem{kim} Kimmel GJ, Locke FL, Altrock PM. Response to CAR-T cell therapy can be explained by ecological cell dynamics and stochastic extinction events. Cold Spring Harb Lab 2019. http://dx.doi.org/10.1101/717074

\bibitem{barros1} Barros LRC, de Jesus Rodrigues B, Almeida RC. CAR-T cell Goes on a Mathematical Model. J Cell Immunol 2020;2(1):31-37. http://dx.doi.org/10.31579/2692-1997/011

\bibitem{mahlbacher} Mahlbacher GE, Reihmer KC, Frieboes HB. Mathematical modeling of tumor-immune cell interactions. J Theor Biol 2019;469:47-60. https://doi.org/10.1016/j.jtbi.2019.03.015

\bibitem{HM} Zhang X, Zhu L, Zhang H, Chen S, Xiao Y. CAR-T Cell Therapy in Hematological Malignancies: Current Opportunities and Challenges. Front. Immunol. 2022;13.  http://dx.doi.org/10.3389/fimmu.2022.927153

\bibitem{HM1}  Freyer CW, Porter DL. Advances in CAR-T Therapy for Hematologic Malignancies. Pharmacotherapy. 2020;40(8):741–755.  http://dx.doi.org/10.1002/phar.2414

\bibitem{ST} Marofi F, Motavalli R, Safonov VA, et al. CAR-T cells in solid tumors: challenges and opportunities. Stem Cell Res Ther. 2021;12(1).  http://dx.doi.org/10.1186/s13287-020-02128-1

\bibitem{ST1} Newick K, O’Brien S, Moon E, Albelda SM. CAR-T Cell Therapy for Solid Tumors. Annu. Rev. Med. 2017;68(1):139–152. http://dx.doi.org/10.1146/annurev-med-062315-120245

\bibitem{eapen} Eapen RS, Williams SG, Macdonald S, Keam SP, Lawrentschuk N, Au L, et al. Neoadjuvant lutetium PSMA, the TIME and immune response in high-risk localized prostate cancer. Nat Rev Urol 2024;1-11. https://doi.org/10.1038/s41585-024-00407-1


\end{thebibliography}
\end{document}